\documentclass[preprint,showpacs,preprintnumbers,amsmath,amssymb]{revtex4}
\usepackage{graphicx}
\usepackage{graphics}
\usepackage{dcolumn}
\usepackage{bm}
\usepackage{amsfonts}
\usepackage{indentfirst}
\usepackage{epsf}
\usepackage{longtable}

 \setlongtables

\begin{document}

\preprint{APS/123-QED}

\title{Information Entropy, Information Distances \\ and
Complexity in Atoms}

\author{K.~Ch.~Chatzisavvas}\email{kchatz@auth.gr}
\author{Ch.~C.~Moustakidis}\email{moustaki@auth.gr}
\author{C.~P.~Panos}\email{chpanos@auth.gr}
\affiliation{
Department of Theoretical Physics,\\ Aristotle University of Thessaloniki,\\
54124 Thessaloniki, Greece}

\date{\today}

\begin{abstract}
Shannon information entropies in position and momentum spaces and
their sum are calculated as functions of $Z$ ($2 \leq Z \leq 54$)
in atoms. Roothaan-Hartree-Fock electron wave functions are used.
The universal property $S=a+b\,\ln{Z}$ is verified. In addition,
we calculate the Kullback-Leibler relative entropy, the
Jensen-Shannon divergence, Onicescu's information energy and a
complexity measure recently proposed. Shell effects at closed
shells atoms are observed. The complexity measure shows local
minima at the closed shells atoms indicating that for the above
atoms complexity decreases with respect to neighboring atoms. It
is seen that complexity fluctuates around an average value,
indicating that the atom cannot grow in complexity as $Z$
increases. Onicescu's information energy is correlated with the
ionization potential. Kullback distance and Jensen-Shannon
distance are employed to compare Roothaan-Hartree-Fock density
distributions with other densities of previous works.
\end{abstract}

\maketitle

\section{Introduction}\label{sec:intro}
Information-theoretic properties are used in recent years for the
study of quantum mechanical systems
\cite{Ohya93,Bialynicki75,Panos97,Massen02,Massen98,Massen01,Garde85,Garde87,Ghosh84,Lalazissis98,
Moustakidis01,Panos01,Panos01b,Massen03,Moustakidis03,Massen05,Psonis,Chatzisavvas05}.
In the present work we carry out a systematic study of Shannon
information $S$, Onicescu information energy $E$, order parameter
$\Omega$ and complexity $\Gamma_{\alpha, \beta}$, in atoms with
$Z=2-54$. In previous work \cite{Massen98} we proposed a universal
property of $S$ for density distributions of nuclei, electrons in
atoms and valence electrons in atomic clusters. This property has
the form
\begin{equation}\label{eq:ini}
    S=a+b\,\ln{N}
\end{equation}
where $N$ is the number of particles of the system and the
parameters $a$ and $b$ depend on the system under consideration.
Recently \cite{Massen02} we have obtained the same form for
systems of correlated bosons in a trap. In the present paper we
employ very accurate spin-independent atomic wave functions
obtained by Bunge et al \cite{Bunge93}, by applying the
Roothaan-Hartree-Fock method (RHF) to calculate analytical self
consistent-field atomic wave functions. Thus we verify the above
relation for atoms, which was obtained in the past
\cite{Garde85,Garde87} employing another set of electron wave
functions. Thus we obtain a framework to be used as a basis for
further work on information-theoretic properties of atoms. An
interesting question is the effect of the electron-electron
correlations on $S$ and relation (\ref{eq:ini}), which could be
answered in future research.

We focus our attention on the problem of similarity index based on
the concept of an information distance. The concept of similarity
is an old one and related to the distinction between two or more
objects \cite{Ho98}. Specifically, in our paper we study two
candidates connected with the concept of similarity or information
distance. The first one is the Kullback-Leibler relative entropy
$K$ (and also the symmetrized Kullback distance $SK$) and the
second one is the Jensen Shannon divergence $J$. Thus we are able
to compare various density distributions obtained using various
models. It turns out that $S$, $SK$ and $J$ measures are useful
for this purpose.

The framework developed in the present work for $S$ together with
$S_{max}$ obtained previously \cite{Garde87} with rigorous
inequalities holding for $S$, $S_r$, $S_k$ enables us to calculate
the so called complexity measure $\Gamma_{\alpha,\beta}$
introduced in \cite{Shiner99}. Our procedure leads to the
interesting result that complexity as function of $Z$ shows shell
effects at closed shells atoms i.e. for the above atoms complexity
decreases with respect to neighboring atoms.

The outline of our paper is the following: In Sec. \ref{sec:2} we
describe measures of information content of a quantum system
together with information distances of two probability
distributions and a complexity measure recently proposed. In Sec.
\ref{sec:3} we present our formalism, while Sec. \ref{sec:4}
contains our numerical results and discussion. Sec. \ref{sec:5} is
a summary of the paper. Finally Sec. \ref{sec:6} contains our
comments on the importance of our results on complexity.

\section{Measures of information content and information
distances}\label{sec:2}

The Shannon information entropy \cite{Shannon48} $S_r$ in
position-space may be defined as
\begin{equation}\label{eq:sr}
    S_r=-\int \rho(\textbf{r})\,\ln{\rho(\textbf{r})}\,d\textbf{r}
\end{equation}
where $\rho(\textbf{r})$ is the electron density distribution
normalized to unity. The corresponding information entropy $S_k$
in the momentum space representation is
\begin{equation}\label{eq:sk}
    S_k=-\int n(\textbf{k})\,\ln{n(\textbf{k})}\,d\textbf{k}
\end{equation}
where $n(\textbf{k})$ is the momentum density distribution
normalized to unity studied in \cite{Garde85,Ho1,Ho2,Ho3}. In
position-space $S_r$ determines the extent of electron
delocalization, since it tends to a maximum as the distribution
flattens out and deviates from this maximum when structure is
introduced in $\rho(\textbf{r})$. In momentum-space a maximum in
$S_k$ corresponds to a delocalized distribution in momentum-space
\cite{Ramirez98}. The total information entropy is given by
\begin{equation}\label{eq:stot}
    S=S_r+S_k
\end{equation}
where $S$, $S_r$ and $S_k$ obey the following rigorous
inequalities \cite{Garde87}

\begin{eqnarray}
S_{r\,min}& \leq S_r & \leq S_{r\,max} \\
S_{k\,min}& \leq S_k & \leq S_{k\,max} \\
S_{min}& \leq S & \leq S_{max}
\end{eqnarray}

The lower and the upper limits can be written, for density
distributions normalized to one
\begin{eqnarray}
S_{r\,min}& = & \frac{3}{2}\,(1+\ln{\pi})-\frac{3}{2}\,\ln{\left(\frac{4}{3}\,T\right)} \nonumber \\
S_{r\,max}& = &
\frac{3}{2}\,(1+\ln{\pi})+\frac{3}{2}\,\ln{\left(\frac{2}{3}\,\langle
r^2 \rangle \right)} \\
\nonumber \\
S_{k\,min}& = &
\frac{3}{2}\,(1+\ln{\pi})-\frac{3}{2}\,\ln{\left(\frac{2}{3}\,\langle
r^2 \rangle \right)} \nonumber \\
S_{k\,max}& = &
\frac{3}{2}\,(1+\ln{\pi})+\frac{3}{2}\,\ln{\left(\frac{4}{3}\,T\right)}
\\
\nonumber \\
S_{min}& = & 3\,(1+\ln{\pi}) \nonumber \\
S_{max}& = &
3\,(1+\ln{\pi})+\frac{3}{2}\,\ln{\left(\frac{8}{9}\,\langle r^2
\rangle \, T\right)}
\end{eqnarray}
where $\langle r^2 \rangle $ is the mean square radius and $T$ is
the kinetic energy.

Another measure of information content of a quantum system is the
concept of information energy $E$ introduced by Onicescu in an
attempt to define a finer measure of dispersion distribution than
that of Shannon information entropy \cite{Onicescu96}. For a
discrete probability distribution $(p_1,p_2,\ldots,p_k)$, $E$ is
defined as
\begin{equation}\label{eq:onidis}
    E=\sum_{i=1}^{k} p_i^2
\end{equation}
which is extended for a continuous density distribution $\rho(x)$
as
\begin{equation}\label{eq:onicont}
    E=\int \rho^2(x)\,dx
\end{equation}
So far, only the mathematical aspects of the concept have been
developed, while the physical aspects have been ignored. A recent
study of $E$ for atomic nuclei has been carried out in
\cite{Moustakidis05}.

The meaning of (\ref{eq:onicont}) can be seen by the following
simple argument: For a Gaussian distribution of mean value $\mu$,
standard deviation $\sigma$ and normalized density
\begin{equation}\label{eq:normdens}
  \rho(x)=\frac{1}{\sqrt{2\pi}\sigma}\, \textrm{exp}
  \left[-\frac{(x-\mu)^2}{2\sigma^2} \right]
\end{equation}
relation (\ref{eq:onicont}) gives
\begin{equation}\label{eq:oninorm}
  E=\frac{1}{2\pi\sigma^2} \int_{-\infty}^{\infty}
  \textrm{exp}
  \left[-\frac{(x-\mu)^2}{\sigma^2}
  \right]\,dx=\frac{1}{2\sigma\sqrt{\pi}}
\end{equation}
$E$ is maximum if one of the $p_i's$ equals 1 and all the others
are equal to zero i.e. $E_{max}=1$, while $E$ is minimum when
$p_1=p_2=\ldots=p_k=\frac{1}{k}$, hence $E_{min}=\frac{1}{k}$
(total disorder). The fact that $E$ becomes minimum for equal
probabilities (total disorder), by analogy with thermodynamics, it
has been called information energy, although it does not have the
dimension of energy \cite{Lepadatu04}.

It is seen from (\ref{eq:oninorm}) that the greater the
information energy, the more concentrated is the probability
distribution, while the information content decreases. Thus one
can define a measure of information content analogous to Shannon's
$S$ by the relation
\begin{equation}\label{}
   O=\frac{1}{E}
\end{equation}
Relation (\ref{eq:onicont}) is extended for a 3-dimensional
spherically symmetric density distribution $\rho(r)$:
\begin{eqnarray}
E_r&=&\int_0^{\infty} \rho^2(r)\,4\pi\,r^2\,dr \nonumber \\
E_k&=&\int_0^{\infty} n^2(k)\,4\pi\,k^2\,dk
\end{eqnarray}
in position and momentum space respectively, where $n(k)$ is the
corresponding density distribution in momentum space.

$E_r$ has dimension of inverse volume, while $E_k$ of volume. Thus
the product $E_r E_k$ is dimensionless and can serve as a measure
of concentration (or information content) of a quantum system. It
is also seen from (\ref{eq:oninorm}) that $E$ increases as
$\sigma$ decreases (or concentration increases) and Shannon's
information entropy (or uncertainty) $S$ decreases. Thus $S$ and
$E$ are reciprocal. In order to be able to compare them, we
redefine the quantity $O$ by
\begin{equation}\label{}
   O=\frac{1}{E_r E_k}
\end{equation}
as a measure of the information content of a quantum system in
both position and momentum spaces.

Landsberg \cite{Landsberg} defined the order parameter $\Omega$
(or disorder $\Delta$) as
\begin{equation}
\Omega=1-\Delta=1-\frac{S}{S_{max}} \label{Landsberg-1}
\end{equation}
where $S$ is the information entropy (actual) of the system and
$S_{max}$ the maximum entropy accessible to the system. Thus the
concepts of entropy and disorder are decoupled and it is possible
for the entropy and order to increase simultaneously. It is noted
that $\Omega=1$ corresponds to perfect order and predictability,
while $\Omega=0$ means complete disorder and randomness.

In \cite{Shiner99} a measure of complexity $\Gamma_{\alpha,\beta}$
was defined of the form
\begin{equation}\label{}
 \Gamma_{\alpha,\beta}=\Delta^{\alpha}\,\Omega^{\beta}=
 \Delta^{\alpha}\,(1-\Delta)^{\beta}=\Omega^{\beta}\,(1-\Omega)^{\alpha}
\end{equation}
which is called the "simple complexity of disorder strength
$\alpha$ and order strength $\beta$". When $\beta=0$ and $\alpha
>0$ "complexity" is an increasing function of "disorder", and we
have a measure of category I (Fig.1 of \cite{Shiner99}). When
$\alpha =0$ and $\beta >0$, "complexity" is an increasing function
of "order" and we have a measure of category III. When both
$\alpha$ and $\beta$ are nonvanishing and positive ($\alpha
>0$, $\beta >0$), "complexity" vanishes at zero "disorder" and
zero "order" and has a maximum of
\begin{equation}\label{}
  (\Gamma_{\alpha,\beta})_{max}=\alpha^{\alpha}\,\beta^{\beta}/(\alpha+\beta)^{(\alpha+\beta)}
\end{equation}
at $\Delta=\alpha/(\alpha+\beta)$ and
$\Omega=\beta/(\alpha+\beta)$. This is complexity of category II
according to \cite{Shiner99}.

Several cases for both $\alpha$ and $\beta$ non-negative are shown
in fig.2 of \cite{Shiner99} where $\Gamma_{\alpha,\beta}$ is
plotted as function of $\Delta$. In the present work we can find
$\Delta=S/S_{max}$ or $\Omega=1-\Delta$ as function of $Z$. Thus
we are able to plot the dependence of $\Gamma_{\alpha,\beta}$ on
the atomic number $Z$.

 The Kullback-Leibler relative information entropy $K$
\cite{Kullback59} for any probability distributions $p_i^{(1)},
p_i^{(2)}$ is defined by
\begin{equation}\label{eq:eq6}
    K(p_i^{(1)},p_i^{(2)})=\sum_i p_i^{(1)}\,\ln{\frac{p_i^{(1)}}{p_i^{(2)}}}
\end{equation}
which for continuous probability distributions is defined as
\begin{equation}\label{eq:eq7}
    K=\int \rho^{(1)}(x)\,\ln{\frac{\rho^{(1)}(x)}{\rho^{(2)}(x)}}\,dx
\end{equation}
(\ref{eq:eq6}), (\ref{eq:eq7}) can be easily extended for
3-dimensional systems. $K$ measures the difference of distance of
$\rho_i^{(1)}$ from the reference (or a priori) distribution
$\rho_i^{(2)}$.

It satisfies: $K\geq 0$ for any distributions $\rho_i^{(1)},
\rho_i^{(2)}$. It is a measure which quantifies the
distinguishability (or distance) of $\rho_i^{(1)}$ from
$\rho_i^{(2)}$, employing a well-known concept in standard
information theory. In other words it describes how close is
$\rho_i^{(1)}$ to $\rho_i^{(2)}$ by carrying out observations or
coin tossing, namely trials $L$ (in the sense described in
\cite{Wooters}).

However, the distance $K$ does not satisfy the triangle inequality
and in addition is i) not symmetric ii) unbounded and iii) not
always well defined \cite{Wooters}. To avoid these difficulties
Rao and Lin \cite{Rao87,Lin91} introduced a symmetrized version of
$K$ \cite{Wooters}, the Jensen-Shannon divergence
\begin{equation}\label{}
  J(\rho^{(1)},\rho^{(2)})=H\left( \frac{\rho^{(1)}+\rho^{(2)}}{2}
  \right)-\frac{1}{2} H(\rho^{(1)})-\frac{1}{2} H(\rho^{(2)})
\end{equation}
where $H(\rho)=-\sum_i \rho_i\,\ln{\rho_i}$ stands for Shannon's
entropy. $J$ is minimum for $\rho^{(1)}=\rho^{(2)}$ and maximum
when $\rho^{(1)}$ and $\rho^{(2)}$ are two distinct distributions,
when $J=\ln{2}$. In our case  $J$ can be easily generalized for
continuous density distributions. For $J$ minimum the two states
represented by $\rho^{(1)}$ and $\rho^{(2)}$ are completely
indistinguishable, while for $J$ maximum they are completely
distinguishable. The amount of distinguishability can be further
examined by using Wooters' criterion \cite{Wooters}. Two
probability distributions $\rho^{(1)}$ and $\rho^{(2)}$ are
distinguishable after $L$ trials $(L\rightarrow \infty)$ if and
only if $\left( J(\rho^{(1)},\rho^{(2)})
\right)^{\frac{1}{2}}>\frac{1}{\sqrt{2L}}$.

The present work is a first step to examine the problem of
comparison of probability distributions (for atoms and various
models) which is an area well developed in statistics, known as
information geometry \cite{Rao87}.

\section{Formalism}\label{sec:3}
The key quantities in our work are the density distribution
$\rho({\bf r})$ and the momentum distribution $n({\bf k})$. In
general the calculation of $\rho({\bf r})$ and $n({\bf k})$
presuppose the knowledge of the one body density matrix $\rho({\bf
r}_1,{\bf r}_1')$ which is defined as

\begin{equation}
\rho({\bf r}_1,{\bf r}_1')=\int \Psi^{*}({\bf r}_1,{\bf
r}_2,\cdots,{\bf r}_Z) \Psi({\bf r}_1',{\bf r}_2,\cdots,{\bf r}_Z)
d{\bf r}_2 \cdots d{\bf r}_Z \label{OBDM-1}
\end{equation}
where $\Psi({\bf r}_1,{\bf r}_2,\cdots,{\bf r}_Z)$ is the wave
function that describes the system under consideration (in the
present work the atom). $\rho(\textbf{r})$ is just the diagonal
part of the $\rho({\bf r}_1,{\bf r}_1')$, i.e.
\[\rho({\bf r}_1)=\rho({\bf r}_1,{\bf r}_1')|_{{\bf r}_1={\bf r}_1'}\]
while $n(\textbf{k})$ is the Fourier transform of the one-body
density matrix i.e.
\[n({\bf k})=\int \rho({\bf r}_1,{\bf r}_1')
\exp[-i{\bf k}({\bf r}_1-{\bf r}_1')] d{\bf r}_1' d{\bf r}_1 \] In
the framework of the Hartree-Fock approximation, which is applied
in the present work, $\Psi({\bf r}_1,{\bf r}_2,\cdots,{\bf r}_Z)$
has the well known form of a Slater determinant (electrons of the
atoms consist a system which obeys the Fermi-Dirac statistics). It
is easy to prove that in that case $\rho({\bf r}_1,{\bf r}_1')$
takes the following simple form
\begin{equation}
\rho({\bf r}_1,{\bf r}_1')=\sum_i \phi_i^{*}({\bf
r}_1)\phi_i^{}({\bf r}_1') \label{OMDM-2-1}
\end{equation}
where $\phi_i({\bf r})$ is the single particle wave function
describing the electrons in an atom. The index $i$ runs  over all
$Z$ single particle states. Now $\rho({\bf r})$ and  $n({\bf k})$
are written as
\begin{equation}
\rho(\textbf{r})=\sum_i \phi_i^{*}({\bf r})\phi_i^{}({\bf r})
\label{PD-1}
\end{equation}
\begin{equation}
n(\textbf{k})=\sum_i {\tilde \phi}_i^{*}({\bf k}) {\tilde
\phi}_i({\bf k}) \label{MD-1}
\end{equation}
where
\begin{equation}
{\tilde \phi}_i({\bf k})=\int \textrm{e}^{-i{\bf k}{\bf r}}
\phi_i^{}({\bf r}) d{\bf r} \label{MWF-1}
\end{equation}
The wave functions  $\phi_i({\bf r})$ and ${\tilde \phi}_i({\bf
k})$ are decomposed in the usual form
\[ \phi_i({\bf r})=\phi_{nlm}({\bf r})=R_{nl}(r) Y_{lm}(\Omega_r) \]
and
\[{\tilde  \phi}_i({\bf k})=
{\tilde \phi} _{nlm}({\bf k})={\tilde R}_{nl}(k) Y_{lm}(\Omega_k)
\]
The radial momentum wave function ${\tilde R}_{nl}(k)$ is related
to the radial wave function in coordinate space through
\begin{equation}
{\tilde R}_{nl}(k)=4 \pi \int_{0}^{\infty} r^2 R_{nl}(r) j_l(kr)
dr \label{Rr-Rk-1}
\end{equation}
where $j_l(kr)$ is a spherical Bessel function.

In the present work we consider very accurate spin-independent
atomic wave functions obtained by  Bunge {\it et al}
\cite{Bunge93} by applying the Roothaan-Hartree-Fock method to
calculate analytical self-consistent-field atomic wave function.
In this approach the radial atomic orbitals $R_{nl}$ are expanded
as a finite superposition of primitive radial functions
\begin{equation}
R_{nl}(r)=\sum_j C_{jnl} S_{jl}(r) \label{}
\end{equation}
where the normalized primitive basis $S_{jl}(r)$ is taken as a
Slater-type orbital set,
\begin{equation}
S_{jl}(r)=N_{jl} r^{n_{jl}-1}e^{-Z_{jl}r} \label{}
\end{equation}
where the normalization factor $N_{jl}$ is given by
\begin{equation}
N_{jl}=(2Z_{jl})^{(n_{jl}+1/2)}/[(2n_{jl})!]^{1/2} \label{}
\end{equation}
and $n_{jl}$ is the principal quantum number, $Z_{jl}$ is the
orbital exponent, and $l$ is the azimuthal quantum number.

In general $\rho({\bf r})$ is not spherically symmetric and
depends in addition on the angle $\theta$. In the present work we
consider $\rho({\bf r})$ averaged spherically (the same holds also
for $n({\bf k})$).

The Shannon information entropy in position-space, momentum space
and in total are given in Eqs. (\ref{eq:sr}), (\ref{eq:sk}) and
(\ref{eq:stot}) respectively.

Another quantity which gives information about the localization
(delocalization) of the atomic systems is the local Shannon
entropy defined (in position- and momentum-space respectively) as
follows
\begin{eqnarray}\label{eq:localshannon}
S_r^{LOC}(\textbf{r})&=&-4 \pi r^2 \rho(\textbf{r}) \ln \rho(\textbf{r}) \nonumber \\
S_k^{LOC}(\textbf{k})&=&-4 \pi k^2 n(\textbf{k}) \ln n(\textbf{k})
\end{eqnarray}

In order to formulate the concept of the similarity or information
distance between two atomic systems with $\rho_A({\bf r})$ and
$\rho_B({\bf r})$ the corresponding density distributions, the
Kullback-Leibler relative entropy is defined as
\begin{equation}
K=\int \rho_A({\bf r}) \ln \frac{\rho_A({\bf r})}{\rho_B({\bf r})}
d{\bf r} \label{K-1}
\end{equation}
which may be interpreted as a measure of deviation of $\rho_A({\bf
r})$ from $\rho_B({\bf r})$. The corresponding symmetrized
Kullback distance $SK$ is
\begin{equation}
SK=\int \rho_A({\bf r}) \ln \frac{\rho_A({\bf r})}{\rho_B({\bf
r})} d{\bf r}+ \int \rho_B({\bf r}) \ln \frac{\rho_B({\bf
r})}{\rho_A({\bf r})} d{\bf r} \label{Id-1}
\end{equation}
The physical meaning of the Kullback distance is very clear. $K$
is equal to zero for two identical species and approaches infinity
as the difference between $\rho_A({\bf r})$ and $\rho_B({\bf r})$
increases. $K$ and $SK$ in momentum space are defined using
$n_A({\bf k})$ and $n_B({\bf k})$ in the same way.

The Jensen-Shannon divergence entropy is defined as follows
\begin{eqnarray}
J&=&-\int \left(\frac{\rho_A({\bf r})+\rho_B({\bf r})}{2} \right)
\ln \left(\frac{\rho_A({\bf r})+\rho_B({\bf r})}{2}
\right) d{\bf r} \nonumber \\
&+&\frac{1}{2}\int \rho_A({\bf r}) \ln \rho_A({\bf r}) d{\bf r}+
\frac{1}{2}\int \rho_B({\bf r}) \ln \rho_B({\bf r}) d{\bf r}
\label{Jensen-1}
\end{eqnarray}
The physical meaning of $J$ is similar to $SK$, while the
definition in momentum space is defined using $n({\bf k})$ in the
same way.

Another aspect of our work is to compare RHF densities with those
of the work \cite{Sagar01} in the framework of the $SK$ and $J$.
More specifically we compare the density distributions of the
electrons originating from the paper of Bunge et.al.
\cite{Bunge93} (used in the present work) with  the
phenomenological one of Sagar et.al. \cite{Sagar01} who employed
the phenomenological form
\begin{equation}
\rho(r)=\frac{(2 I_1)^{3/2}}{\pi} e^{-2(2I_1)^{1/2}r}
\label{d-sag-1}
\end{equation}
%
%
where $I_1$ is the first ionization potential of the system. We
also compare RHF densities with those of the well known
Thomas-Fermi model. For the density distribution we use the simple
form obtained by Sommerfeld \cite{Sommerfeld32} where
\begin{equation}
\rho(r)=C_{norm}\frac{2^{3/2} Z^{3/2}}{3 \pi^2 r^{3/2}}
\left(1+\left(\frac{r}{\mu \alpha}\right)^d
\right)^{-\frac{3}{2}c} \label{TF-DD}
\end{equation}
where $\alpha=12^{2/3}$, $d=0.772$, $c=3.886$ and
$\mu=0.885341/Z^{1/3}$. The normalization constant $C_{norm}$ is
calculated from the normalization condition $\int \rho(\textbf{r})
d{\bf r}=1$

\section{Numerical results and discussion}\label{sec:4}

\begin{figure}
 \centering
 \includegraphics[height=7.0cm,width=6.0cm]{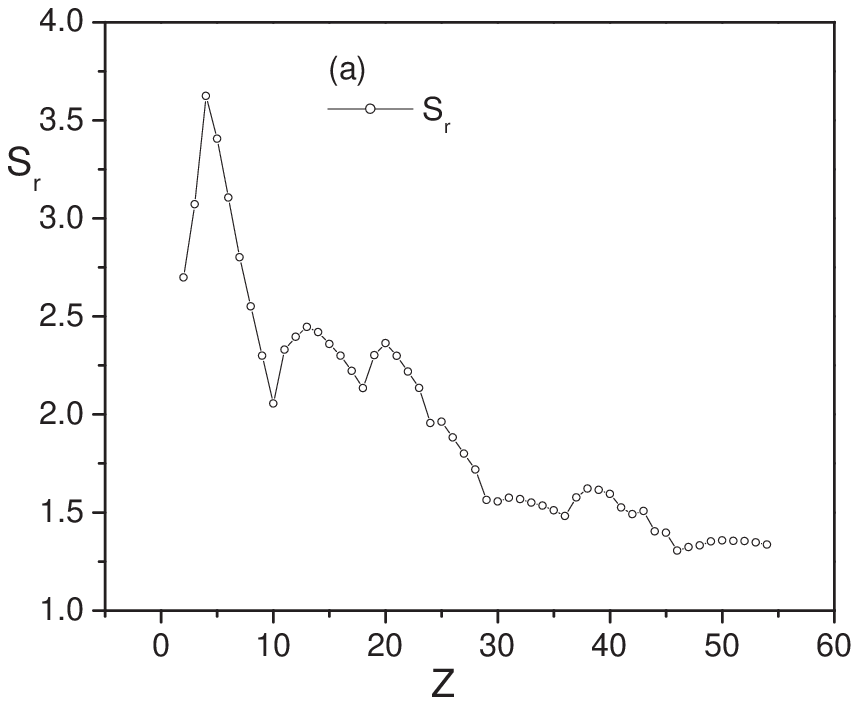}
 \\
 \includegraphics[height=7.0cm,width=6.0cm]{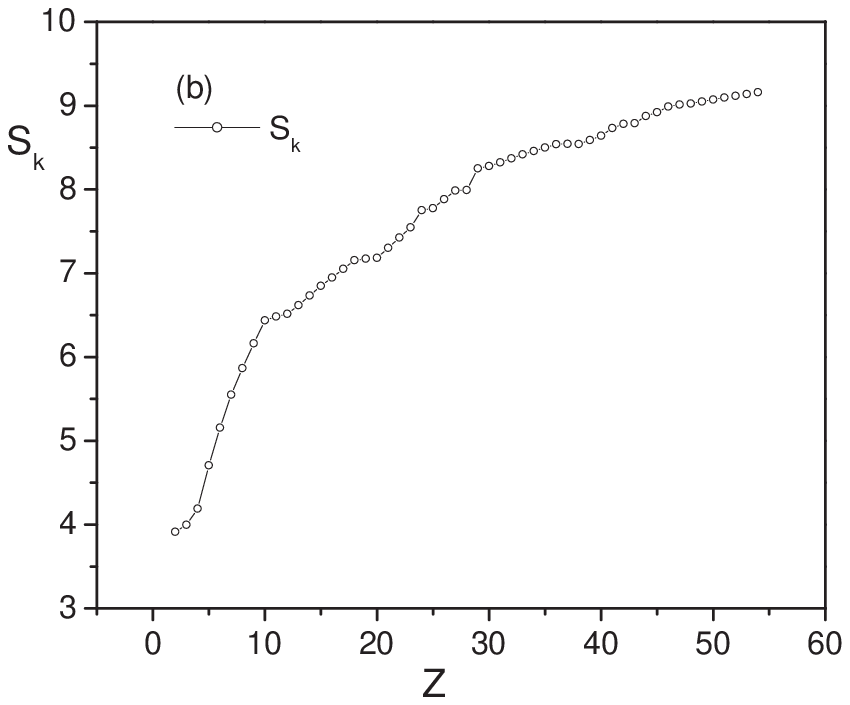}
 \caption{Shannon information entropy (a) in coordinate-space $S_r$
and (b) in momentum-space $S_k$, as a function of the electron
number $Z$.\label{fig:1}}
 \end{figure}

In Fig.~\ref{fig:1} (a) and (b)  we plot the Shannon information
entropy both in coordinate-space ($S_r$) and momentum-space
($S_k$) as functions of the electron number $Z$. In $S_r$ coexist
an average increasing behavior and also an obvious shell effect
structure around fully filled shells, such as He, Ne, Ar, Kr where
there are minima of the curve $S_r(Z)$. The physical meaning of
that behavior is that $\rho(r)$ for these atoms is the most
compact one when compared to the neighboring atoms. The values of
$S_k$ (Fig.~\ref{fig:1} (b)) show a monotonic increase with $Z$.
However there is also in the behavior of $S_k$ a local shell
effect.

\begin{figure}
 \centering
 \includegraphics[height=7.0cm,width=6.0cm]{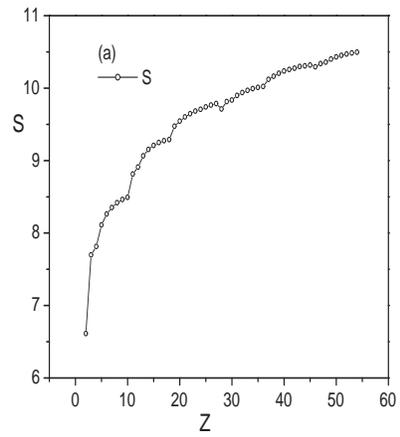}
 \\
 \includegraphics[height=7.0cm,width=6.0cm]{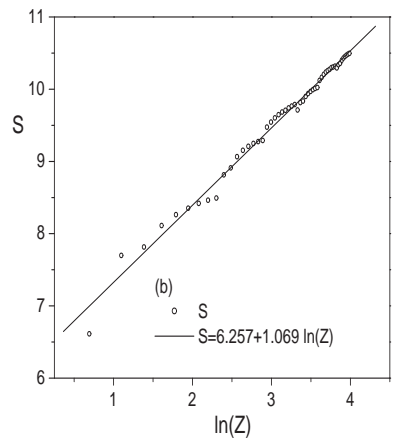}
 \caption{(a)
Total Shannon Information entropy $S$. (b) $S$ as a function of
$\ln{Z}$ and linear fit.\label{fig:2}}
\end{figure}

In Fig.~\ref{fig:2} (a) we plot the total Shannon information
entropy $S$. $S$ is a strictly monotonic increasing function of
$Z$ with only two exceptions for Ni and Pd. These exceptions are
due to the fact that $S_r$ and $S_k$ depend on the arrangement of
the electrons in shells. There is a delicate balance between $S_r$
and $S_k$ resulting in the general rule that $S=S_r+S_k$ is a
monotonic increasing function of $Z$ except in Ni and Pd where the
electron arrangement in shells is such that the decrease of the
value of $S_r$ cannot be balanced by a corresponding increase of
$S_k$. Thus the monotonicity of $S$ is slightly violated. A shell
effect is also obvious in the behavior of $S$ i.e. minima at
closed shells. Fig.~\ref{fig:2} (b) illustrates the trend of $S$
as a function of $\ln{Z}$. The best linear fit is also plotted in
the same figure where $S=6.257+1.069\,\ln{Z}$. It is noted that
this result is not new but has been already obtained using other
wave functions in \cite{Garde85,Garde87}. In the present work we
verify this result with RHF electron wave functions \cite{Bunge93}
and we employ this framework for new calculations.

\begin{figure}
 \centering
 \includegraphics[height=5.5cm,width=6.0cm]{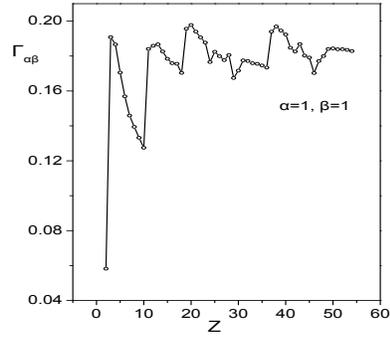}
 \\
 \includegraphics[height=5.5cm,width=6.0cm]{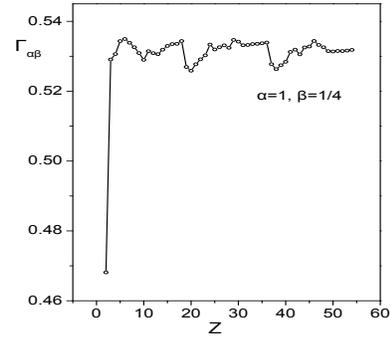}
 \\
 \includegraphics[height=5.5cm,width=6.0cm]{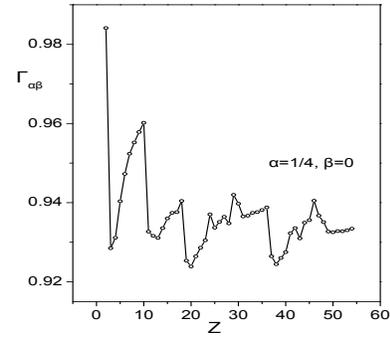}
 \\
 \includegraphics[height=5.5cm,width=6.0cm]{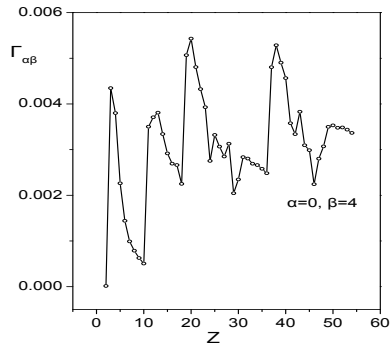}
 \caption{Complexity measure $\Gamma_{\alpha,\beta}$ for
 various sets of parameters $\alpha$ and $\beta$ as a function of $Z$.\label{fig:3}}
\end{figure}

In Fig.~\ref{fig:3} (a), (b), (c), (d) we plot the complexity
measure $\Gamma_{\alpha,\beta}$ in atoms for various values of
parameters $\alpha$ and $\beta$. It is seen that for all sets of
$\alpha$ and $\beta$, $\Gamma_{\alpha,\beta}$ shows qualitatively
the same trend as function of $Z$ i.e. it fluctuates around an
average value and shows local minima for atoms with closed shells.
These results compare favorably with intuition i.e. complexity is
less at closed shells which can be considered more compact than
neighboring nuclei and consequently less complex. It is noted that
this result comes from a procedure which is not trivial i.e. first
we calculate $\rho({\bf r})$ and $n({\bf k})$, second we find
$S=S_r+S_k$ from the Shannon definition and $S_{max}$ employing
rigorous inequalities and third we obtain the complexity measure
introduced in \cite{Shiner99}.

\begin{figure}
 \centering
 \includegraphics[height=7.0cm,width=6.0cm]{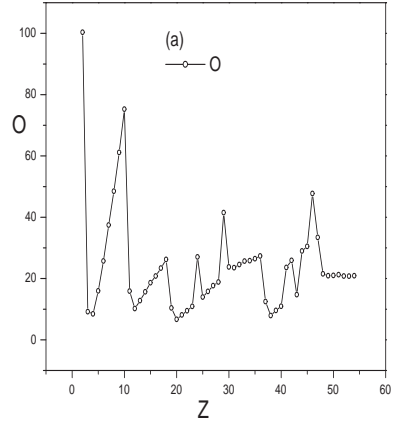}
 \\
 \includegraphics[height=7.0cm,width=6.0cm]{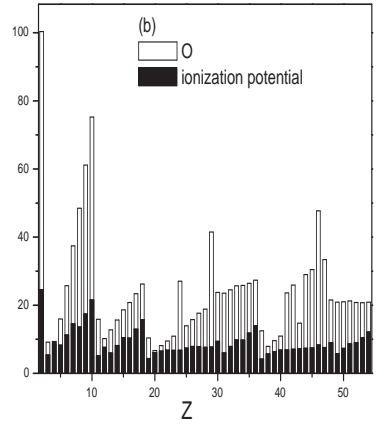}
 \caption{(a)
Onicescu information content $O$ versus $Z$, (b) Onicescu
information content $O$ and ionization potential $I_1$ (in Hartree
units) versus $Z$.\label{fig:4}}
\end{figure}

Fig.~\ref{fig:4} (a) displays the values of the Onicescu
information content $O$ versus $Z$. The first three maxima
correspond to the fully closed shells (He, Ne and Ar) where in the
case of the next closed shell (Kr) a local maximum exists. In
Fig.~\ref{fig:4} (b) we plot in the same footing the Onicescu
information content $O$ and the ionization potential $I_1$. It is
indicated that $O$ and $I_1$ are correlated in the sense that
there is a similarity in the trend of values of $O$ and $I_1$ as
functions of $Z$. This similarity is more obvious in regions of
small $Z$ where linear relations $O=a+b\,\ln{Z}$ can be extracted
for regions Li-Ne and Na-Ar. However, it seems that, there is no
universal relation between them. There are many entropic measures
of spread of probability densities e.g. $S$, $E$ e.t.c. but
researchers prefer $S$ because of its unique properties, while $E$
was introduced by Onicescu as a sensitive measure of information.
However, $S$ and $E$ are different functionals of the density and
their relation is difficult to find.

\begin{figure}
 \centering
 \includegraphics[height=7.0cm,width=6.0cm]{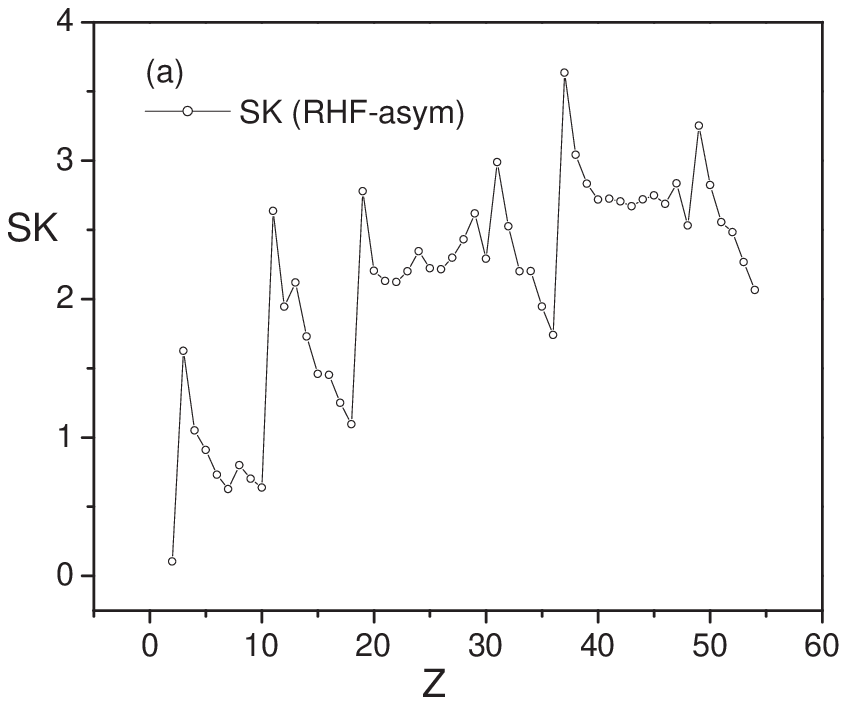}
 \\
 \includegraphics[height=7.0cm,width=6.0cm]{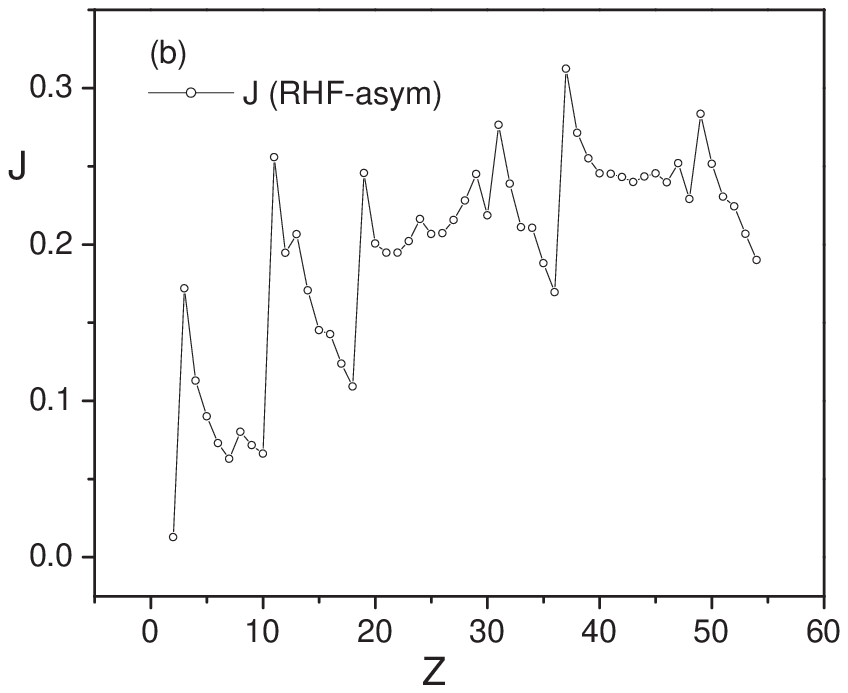}
  \\
 \includegraphics[height=7.0cm,width=6.0cm]{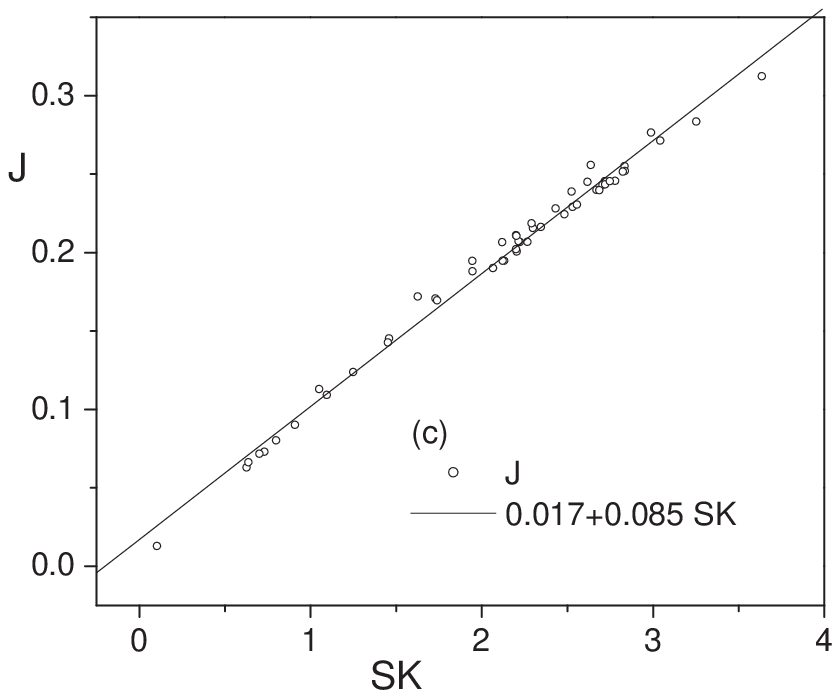}
 \caption{(a)
Symmetrized Kullback distance between the RHF density distribution
and the approximate (asymptotic) one (Eq.~(\ref{d-sag-1})) (b)
corresponding Jensen-Shannon divergence entropy versus $Z$ (c)
Jensen-Shannon divergence versus the symmetrized Kullback
distance.\label{fig:5}}
\end{figure}

In Fig.~\ref{fig:5} (a) we display the symmetrized Kullback
distance between the RHF density distributions and the approximate
one given by the relation (\ref{d-sag-1}). It is obvious that the
symmetrized Kullback distance $SK$ becomes minimum in the case of
the fully closed shells atoms. This physically means that the
approximate $\rho({\bf r})$ (\ref{d-sag-1}) works better in closed
shell atoms than the open shells. In addition there is an
increasing trend with $Z$. That means that in general $\rho({\bf
r})$ (\ref{d-sag-1}) fails to describe the structure of heavier
atoms. Fig.~\ref{fig:5} (b) displays the Jensen-Shannon divergence
entropy $J$ as a function of $Z$. The behavior is almost similar
to $SK$ and the comments are the same. In Fig.~\ref{fig:5} (c) we
display $J$ versus $SK$. There is a linear relation of the two
information distances.

\begin{figure}
 \centering
 \includegraphics[height=7.0cm,width=6.0cm]{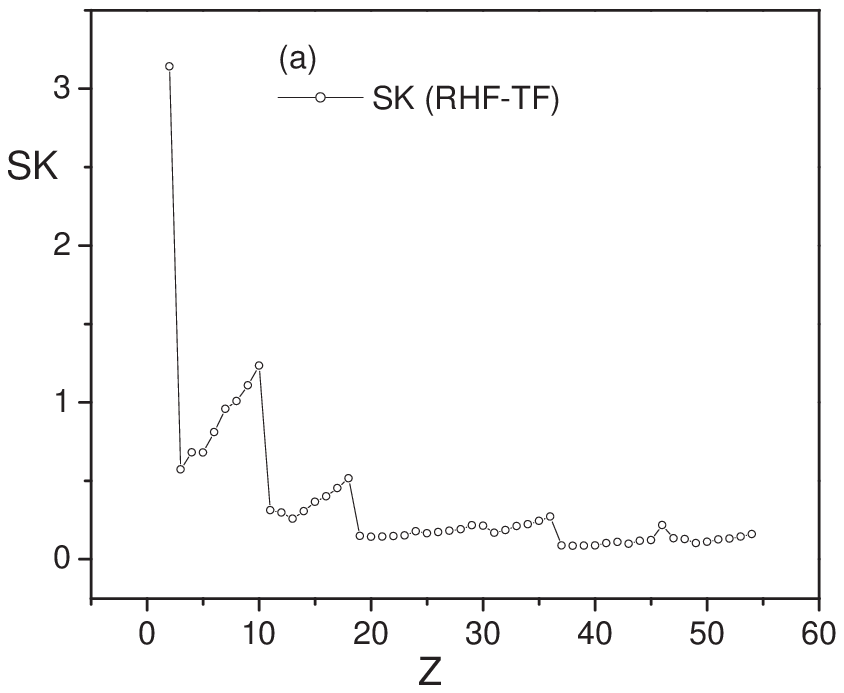}
 \\
 \includegraphics[height=7.0cm,width=6.0cm]{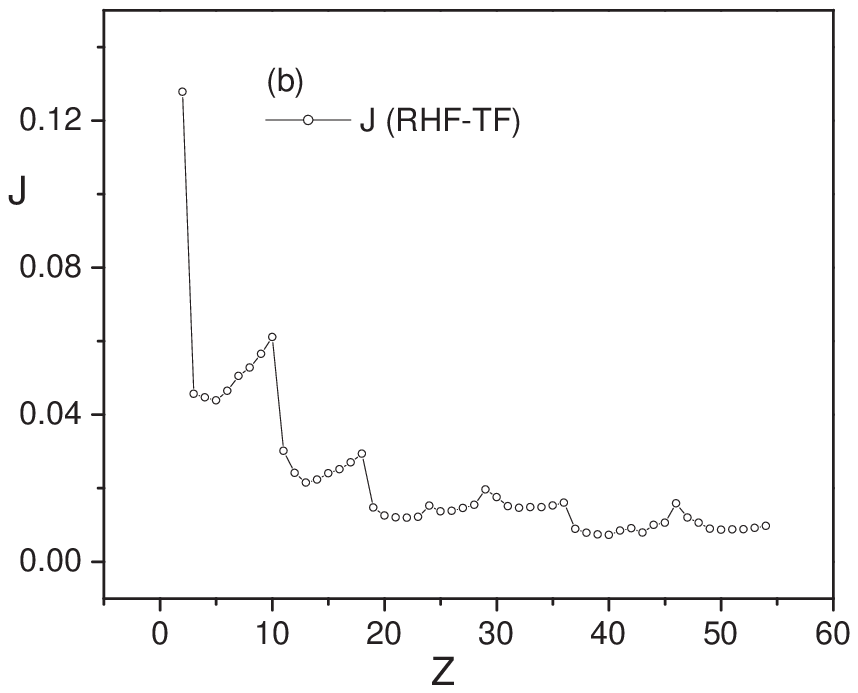}
  \\
 \includegraphics[height=7.0cm,width=6.0cm]{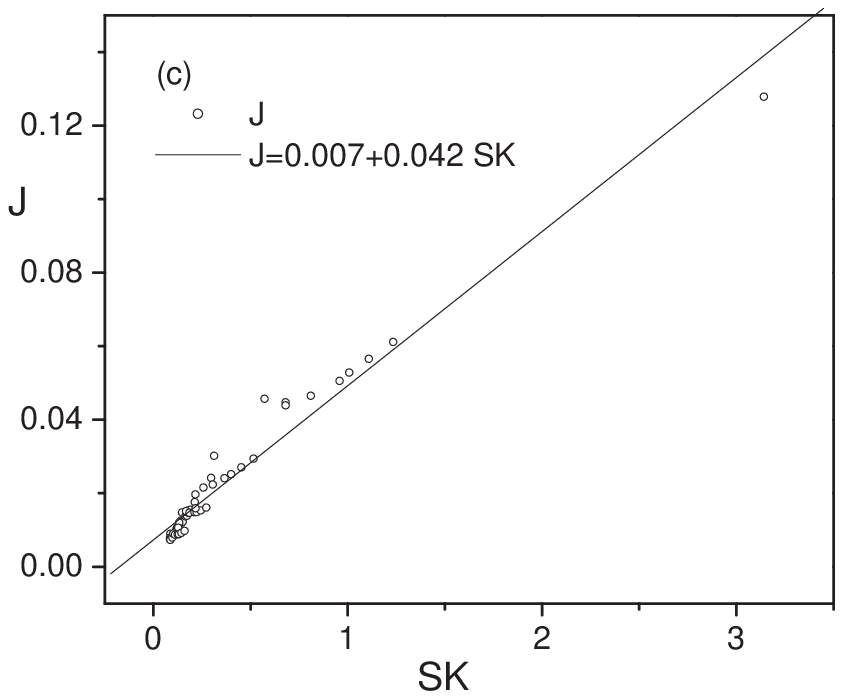}
 \caption{(a)
Symmetrized Kullback distance between the RHF density distribution
and the Thomas-Fermi density distribution as functions of $Z$ (b)
corresponding Jensen-Shannon divergence entropy versus $Z$ (c)
Jensen-Shannon divergence versus the symmetrized Kullback
distance.\label{fig:6}}
\end{figure}

In Fig.~\ref{fig:6} (a) we display $SK$ as an information distance
between the RHF density distribution and the Thomas-Fermi density
distribution given by the relation (\ref{TF-DD}). It is seen that
the values of $SK$ decrease with $Z$. This is expected because the
Thomas-Fermi approximation works well in heavier atoms. So the two
distributions are closer in heavier atoms than in the light ones.
The local maxima of the values of $SK$ correspond to the closed
shell atoms. This deviation is due to the absence of shell effect
character of the Thomas-Fermi distribution compared to the
realistic  RHF density distribution. Fig.~\ref{fig:6} (b) displays
$J$ as a function of $Z$. The behavior is almost similar to $SK$.
In Fig.~\ref{fig:6} (c) we display $J$ versus $SK$.

\begin{figure}
 \centering
 \includegraphics[height=7.0cm,width=6.0cm]{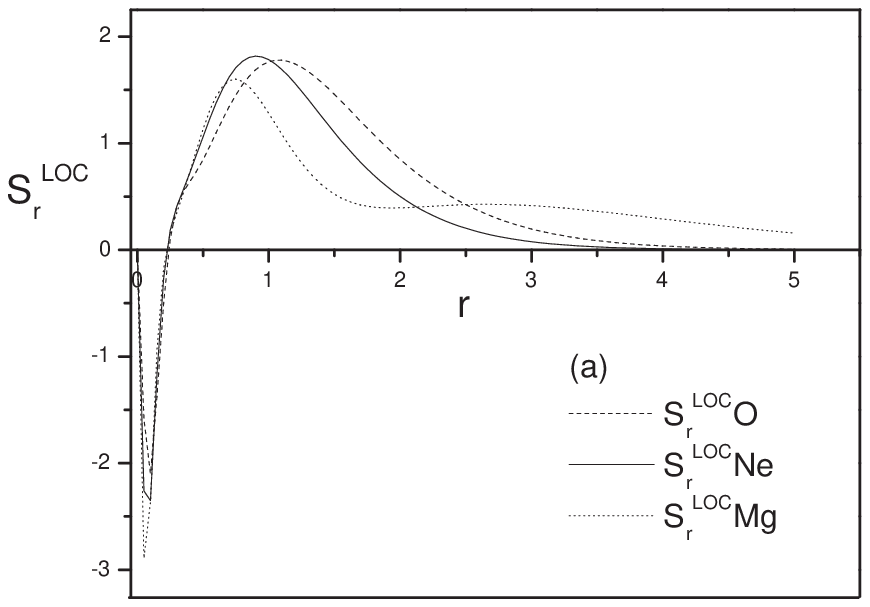}
 \\
 \includegraphics[height=7.0cm,width=6.0cm]{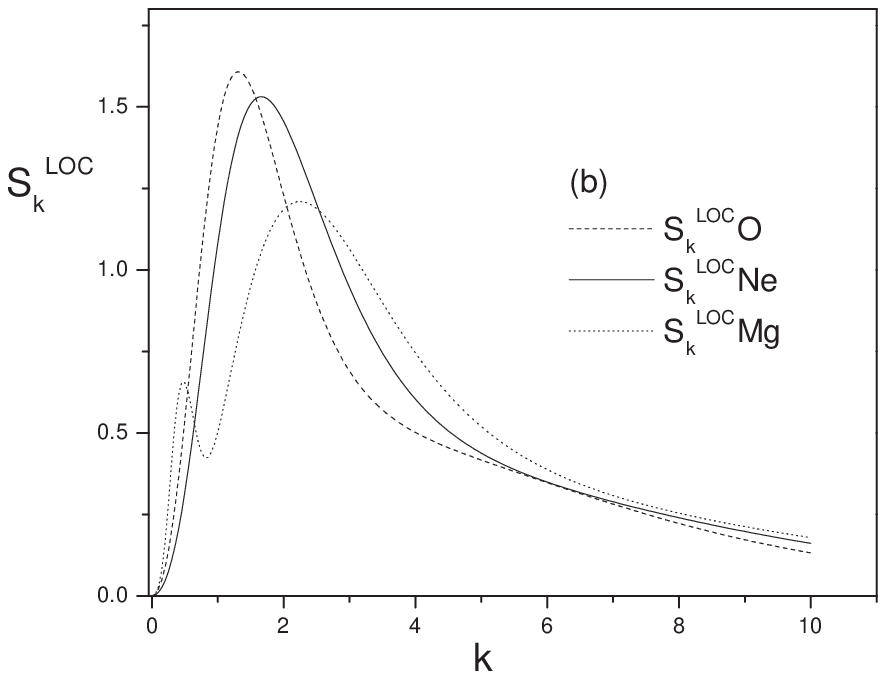}
 \caption{Local
Shannon information entropy (a) in coordinate-space $S_r^{LOC}$
and (b) in momentum-space $S_k^{LOC}$, as a function of the
electron number $Z$.\label{fig:7}}
\end{figure}

The local Shannon entropy (\ref{eq:localshannon}) both in
position- and momentum-space is presented in Fig.~\ref{fig:7} for
various atoms. As it is pointed out in \cite{Guevara} in systems
with density $\rho(\textbf{r})>1$ near the nucleus the local
Shannon entropy in position-space $S_r^{LOC}$ will be negative,
thus the contribution to the integral from this region will serve
to lower its value (localization) while in region where
$\rho(\textbf{r})<1$, such as the valence, contribution to the
Shannon entropy in position-space will be positive which leads to
delocalization. In contrast the local Shannon entropy in
momentum-space $S_k^{LOC}$ is always positive. This is due to the
fact that $n(\textbf{k})<1$ for all $k$. So, there is no negative
contribution to the Shannon entropy in momentum-space.

It is worth, discussing the behavior of the momentum distribution
and as a consequence the local Shannon entropy, to mention that
the precise knowledge of the electron momentum distribution is
important for atoms used as a dark matter or neutrino detectors.
In such a kind of experiments the single particle wave function in
momentum space or in general the momentum density of the electrons
are the main ingredient of the relative cross sections. The trend
(localization, delocalization, etc) of the electron wave function
or momentum distribution affect considerably the values of the
cross sections, especially in experiments where the production of
electrons in neutralino-nucleus or neutrino-nucleus collisions are
investigated \cite{Gounaris,Vergados05,Moustakidis}.

Finally in Table~\ref{tab:1} we tabulate for the sake of reference
the quantities $S_r$, $S_k$, $S$, $S_{max}$, $\Omega$, and $O$ for
each atom as functions of $Z$. We include the results for Hydrogen
which are known exactly i.e. $S_r=3+\ln{\pi}$ and
$S_k=\ln{32\pi^2}-\frac{10}{3}$.

\section{Summary}\label{sec:5}

In previous works the universal relation $S=a+b\,\ln{N}$ was
proposed for the information entropy $S$ as function of the number
of particles $N$ in atoms, nuclei, atomic clusters and correlated
bosons in a trap i.e. systems of various sizes, with various
interactions, obeying different statistics (fermions and bosons).
In this paper we verify the above relation employing RHF electron
density distributions for atoms. Thus we construct a basis in
order to study some information-theoretic properties of atoms.
Specifically, we calculate the symmetrized Kullback-Leibler
relative entropy $SK$ and the Jensen-Shannon divergence $J$ which
serve as measures of information distance of probabilities
distributions and are useful to compare electron distributions
according to various models. Two examples are given. We compare
RHF density distributions first with an asymptotic density
depending on the ionization potential and second with the
well-known Thomas Fermi approximation.

We also obtain Onicescu's information energy $E$ and its
corresponding information measure, which correlates with the
ionization potential. Finally, we calculate a recently proposed
complexity measure $\Gamma_{\alpha,\beta}$ inspired by Landsberg's
order parameter $\Omega$. It turns out that the function
$\Gamma_{\alpha,\beta}(Z)$ shows the interesting feature that for
closed shells atoms is smaller than neighboring ones. This
indicates that closed shells atoms are less complex than
neighboring ones, which compares favorably with expectations
according to intuition.

\section{Final Comments}\label{sec:6}

There is a long debate in the literature on order, disorder,
complexity and organization for physical, biological and other
systems. A generally accepted quantitative definition of
complexity does not exist so far. More work has been carried out
in classical systems and much less in quantum ones. Our aim is to
contribute in this debate examining complexity in atoms (for the
first time to our knowledge) adopting a particular definition of
complexity measure according to  Shiner, Davison and Landsberg
\cite{Shiner99}. The reason for our choice is that we can easily
calculate it for atoms (and we intend to calculate it for other
quantum systems as well), because our previous experience with
information entropy facilitates this. More work is needed in the
future by using the so called statistical measure of complexity of
Lopez-Ruiz, Manchini and Calbet \cite{Lopez95}

In \cite{Landsberg98} the authors studied disorder and complexity
in an ideal Fermi gas of electrons. They observed that for a small
number of electrons Landsberg's order parameter $\Omega$ is small,
while $\Omega$ increases as one pumps electrons into the system
and the energy levels fill up. This result is in a way
counter-intuitive and indicates that as particles are added in a
quantum-mechanical system, the system becomes more ordered. This
result was further corroborated in \cite{Panos01c} by calculating
$\Omega$ as function of the number of particles $N$ for realistic
quantum systems i.e. atomic nuclei and atomic clusters. In
\cite{Massen02} $\Omega(N)$ was obtained for bosonic systems as
well (correlated atoms in a trap). All cases show the same trend
for $\Omega(N)$ i.e. $\Omega$ is an increasing function of $N$.
However, if one is interested how complex (or organized) is a
system, the information entropy $S$ (sometimes used as a measure
of disorder) or the order parameter $\Omega$ (or disorder
$\Delta=1-\Omega$) are not suitable measures of complexity or
organization. We note that these terms are interrelated from a
semantic point of view. $\Gamma_{\alpha,\beta}$ has the advantage
of being a convex measure of order (or disorder) i.e. it vanishes
for highly ordered and disordered systems. In other words, it
satisfies the "one-hump" criterion for statistical complexity
measures. Thus $\Gamma_{\alpha,\beta}$ is obtained by multiplying
a measure of "order" by a measure of "disorder". A simple example
is a perfect crystal which has perfect order and an ideal gas with
complete disorder. Both have zero complexity and fit well with the
definition of $\Gamma_{\alpha,\beta}$.

In the present paper we find that complexity is less at closed
shell atoms. This satisfies our intuition, at least does not
contradict common sense and indicates that our procedure from
electron densities to information entropy and maximum entropy to
complexity measure $\Gamma_{\alpha,\beta}$ is reasonable. The fact
that $\Gamma_{\alpha,\beta}$ fluctuates around an average value is
new and interesting, because it shows that complexity of atoms
does not increase as the atomic number $Z$ increases. In other
words, as one pumps electrons into the atom, the atom has not the
ability to grow in complexity. The question is open what happens
if atoms form molecules, molecules form more complex systems
e.t.c. We mention that the question whether physical or biological
systems have the ability for organized complexity without the
intervention of an external factor or agent is a hot subject in
the community of scientists interested in complexity and can be
extrapolated even in philosophical questions.

\section{Acknowledgments}
The work of K.~Ch.~Chatzisavvas and C.~P.Panos was supported by
Herakleitos Research Scholarships (21866) of
$\textrm{E}\Pi\textrm{EAEK}$ and the European Union while the work
of Ch.~C.~Moustakidis was supported by the Pythagoras II Research
project (80861) of $\textrm{E}\Pi\textrm{EAEK}$ and the European
Union. The authors would like to thank Prof. C.~F.~Bunge for
kindly providing the data for the atomic wave functions, Prof.
S.~E.~Massen for useful comments on the manuscript and also Prof
R.~P.~Sagar for valuable comments and correspondence.

\clearpage
\newpage


\clearpage
\newpage


\begin{longtable*}[c]{l l c c r@{.}l r@{.}l c r@{.}l l}
\caption{\label{tab:1}The values of various quantities in our
systematic study.} \\
$Z$ & Atom &$S_r$ &$S_k$& \multicolumn{2}{c}{$S$} &
\multicolumn{2}{c}{$S_{max}$} & $\Omega$ & \multicolumn{2}{c}{$O$} \\
\hline \hline
1& H & 4.14473 & 2.42186 &6&56659 & 7&9054 &0.21113 &120&26700 \\
2 & He &2.69851 &3.91342 &6&61193 & 7&0493
&0.06204 &100&36100 \\
3 & Li  &3.07144 &3.99682 &7&69826 &
10&3578 &0.25677 &9&15713 \\
4 & Be &3.62386 &4.19019 &7&81405  &10&3950  &0.24829
&8&45434 \\
5 & B &3.40545 &4.70590  &8&11135 & 10&3738 &0.21810
&15&96530 \\
6 & C &3.10602 &5.15658 &8&26260  &10&2624 &0.19492
&25&71210 \\
7 & N &2.80169 &5.54934 &8&35103  &10&1520  &0.17740
&37&43200 \\
8 & O &2.55054 &5.86737 &8&41791 &10&1113  &0.16747
&48&48340 \\
9 &F &2.29883 &6.16333 &8&46215 & 10&0533  &0.15827
&61&14500 \\
10 &Ne &2.05514 &6.43707 &8&49221 & 9&9908  & 0.14999 &75&24470
\\
11 &Na & 2.33009 &6.48310  &8&81319 & 11&6463 &0.24326
&15&86900 \\
12 &Mg &2.39540  &6.51440  &8&91038  &11&8296  &0.24677
&10&19480 \\
13 &Al &2.44569 &6.61928 &9&06497  &12&0615  &0.24843
&12&76590 \\
14 &Si & 2.41914 &6.73380  &9&15294  &12&0500  &0.24042
&15&63600 \\
15 &P & 2.35903 &6.84865 &9&20767  &11&9954  &0.23240
&18&62490 \\
16 &S &2.29932 &6.94939 &9&24871 & 11&9769  &0.22779
&20&80380 \\
17 &Cl &2.22174 &7.05243 &9&27418  &11&9315  &0.22716
&23&36830 \\
18 &Ar &2.13383 &7.15541 &9&28924  &11&8758  &0.21780 &26&21000 \\
19 &K  &2.30177 &7.17242 &9&47419 & 12&9220 &0.26682
&10&37540 \\
20 &Ca &2.36309 &7.18250  &9&54334 & 13&0994  &0.27147 &6&66998
\\
21 &Sc &2.29814 &7.30329 &9&60143 & 13&0334  &0.26332
&8&12090 \\
22 &Ti &2.21855 &7.42693 &9&64548 &12&9721 &0.25644
&9&48793 \\
23 &V &2.13512 &7.54717 &9&68229  &12&9160  &0.25036
&10&91540 \\
24 &Cr &1.95589 &7.75135 &9&70724 &12&5913   &0.22905
&27&02960 \\
25 &Mn &1.96257 &7.77688 &9&73945 &12&8162  &0.24007 &13&94220 \\
26 &Fe  &1.88213 &7.88265 &9&76478 &12&7697  &0.23532 &15&78050
\\
27 &Co  &1.80001 &7.98612 &9&78613 &12&7264  &0.23104
&17&65210 \\
28 &Ni &1.71826 &7.99294 &9&71120  &12&7198 &0.23653
&18&83670 \\
29 &Cu &1.56325 &8.25076 &9&81401 &12&4646  &0.21265
&41&47510 \\
30 &Zn &1.55625 &8.27867 &9&83493 &12&6106  &0.22011
&23&74300 \\
31 &Ga &1.57444 &8.32388 &9&89832 &12&8676  &0.23076
&23&52780 \\
32 &Ge &1.56746 &8.37150  &9&93896 &12&9098  &0.23012
&24&53250 \\
33 &As  &1.54980  &8.41828 &9&96808 &12&9082  &0.22777 &25&68280
\\  34 &Se &1.53425 &8.45851 &9&99276 &12&9298  &0.22715
&25&81220 \\
35 &Br &1.51064 &8.49958 &10&01020  &12&9230 &0.22540
&26&44050 \\
36 &Kr &1.48146 &8.54092 &10&02240 &12&9026  &0.22323 &27&33700
\\37 &Rb &1.57626 &8.54430 &10&12060 &13&7376 &0.26329
&12&45970 \\
38 &Sr &1.62144 &8.54228 &10&16370 &13&9159  &0.26963
&7&91895 \\
39 &Y &1.61438 &8.59046 &10&20480 &13&8772  &0.26463
&9&57134 \\
40 &Zr &1.59462 &8.64054 &10&23520 &13&8304  &0.25995
&10&94010 \\
41 &Nb &1.52486 &8.73236 &10&25720 &13&5780  &0.24457
&23&59820 \\
42 &Mo &1.49117 &8.78233 &10&27350 &13&5247  &0.24039
&25&89550 \\
43 &Tc &1.50762 &8.79074 &10&29840 &13&7087  &0.24877
&14&73730 \\
44 &Ru &1.40305 &8.87561 &10&30610 &13&4876  &0.23588
&28&97120 \\
45 &Rh &1.39635 &8.92206 &10&31840 &13&4670  &0.23380
&30&46670 \\
46 &Pd &1.30482 &8.98908 &10&29390 &13&1570  &0.21761
&47&72420 \\
47 &Ag &1.32346 &9.01391 &10&33740 &13&4264  &0.15816
&33&39990 \\
48 &Cd &1.33132 &9.02613 &10&35740 &13&5464  &0.23541
&21&51740 \\
49 &In &1.35191 &9.04892 &10&40080 &13&7432  &0.24320
&20&89160 \\
50 &Sn &1.35701 &9.07295 &10&43000 &13&7921  &0.24377
&21&00720 \\
51 &Sb &1.35483 &9.09683 &10&45170 &13&8039  &0.24285
&21&24880 \\
52 &Te &1.35345 &9.11719 &10&47060 &13&8313  &0.24297
&20&80400 \\
53 &I &1.34658 &9.13832 &10&48490  &13&8349  &0.24214
&20&74270 \\
54 &Xe &1.33582 &9.16022 &10&49600 &13&8264  &0.24087
&20&89380\\
\hline \hline
\end{longtable*}

\clearpage
\newpage


\clearpage
\newpage

\clearpage
\newpage

\clearpage

\clearpage
\newpage

\clearpage
\newpage

\clearpage
\newpage

\clearpage
\newpage


\begin{thebibliography}{qq}
\bibitem{Ohya93}
 M.~Ohya,  and D.~Petz, \emph{Quantum Entropy and Its Use} (
 Springer-Verlag, Berlin; New York, 1993).

\bibitem{Bialynicki75}
 I.~Bialynicki-Birula, and J.~Mycielski, Commun. Math. Phys.
 \textbf{44}, 129 (1975).

\bibitem{Panos97}
 C.~P.~Panos, and S.~E.~Massen, Int. J. Mod. Phys. E
 \textbf{6}, 497 (1997).

\bibitem{Massen02}
 S.~E.~Massen, Ch.~C.~Moustakidis, and C.~P.~Panos, Phys. Lett. A
 \textbf{64}, 131 (2002).

\bibitem{Massen98}
S.~E.~Massen, and C.~P.~Panos, Phys. Lett. A \textbf{246}, 530
(1998).

\bibitem{Massen01}
 S.~E.~Massen, and C.~P.~Panos, Phys. Lett. A \textbf{280}, 65 (2001).

\bibitem{Garde85}
 S.~R.~Gadre, S.~B.~Sears, S.~J.~Chakravorty, and R.~D.~Bendale, Phys. Rev. A \textbf{32}, 2602 (1985).

\bibitem{Garde87}
 S.~R.~Gadre, and R.~D.~Bendale, Phys. Rev. A \textbf{36}, 1932 (1987).

\bibitem{Ghosh84}
 S.~K.~Ghosh, M.~Berkowitz, and R.~G.~Parr, Proc. Natl. Acad.
 Sc. USA \textbf{81}, 8028 (1984).

\bibitem{Lalazissis98}
 G.~A.~Lalazissis, S.~E.~Massen, C.~P.~Panos, and S.~S. Dimitrova,
 Int. J. Mod. Phys. E \textbf{7}, 485 (1998).

\bibitem{Moustakidis01}
 Ch.~C.~Moustakidis, S.~E.~Massen, C.~P.~Panos, M.~E.~Grypeos,
 and A.~N.~Antonov, Phys. Rev. C \textbf{64}, 014314 (2001).

\bibitem{Panos01}
 C.~P.~Panos, S.~E.~Massen, and C.~G.~Koutroulos, Phys. Rev. C
 \textbf{63}, 064307 (2001).

\bibitem{Panos01b}
 C.~P.~Panos, Phys. Lett. A \textbf{289}, 287 (2001).

\bibitem{Massen03}
 S.~E.~Massen, Phys. Rev. C \textbf{67}, 014314 (2003).

\bibitem{Moustakidis03} Ch.~C.~Moustakidis, and S.~E.~Massen, Phys. Rev. B {\bf 71}, 045102 (2003).

\bibitem{Psonis} S.~E.~Massen, V.~P.~Psonis, and A.~N.~Antonov,
e-print nucl-th/0502047.

\bibitem{Massen05}
 S.~E.~Massen, Ch.~C.~Moustakidis, and C.~P.~Panos, \emph{Focus on Boson
 Research}, edited by A.~V.~Ling, (Nova Publishers, in press).

\bibitem{Chatzisavvas05}
 K.~Ch.~Chatzisavvas, and C.~P.~Panos, Int. J. Mod Phys. E
 (to be published).

\bibitem{Bunge93} C.~F.~Bunge, J.~A.~Barrientos, and A.~V.~Bunge, At. Data
Nucl. Data Tables {\bf 53}, 113 (1993).

\bibitem{Ho98} M.~H$\hat{\textrm{o}}$, V.~H.~Smith,Jr., D.~F.~Weaver, C.~Gatti, R.~P.~Sagar,
and R.~O.~Esquivel, J. Chem. Phys. {\bf 108}, 5469 (1998).

\bibitem{Shiner99} J.~S.Shiner, M.~Davison, and P.~T.~Landsberg,
Phys. Rev. E \textbf{59}, 1459 (1999).

\bibitem{Shannon48} C.~E.~Shannon, Bell Syst. Tech. J.
\textbf{27}, 379 (1948).

\bibitem{Ho1} M.~H$\hat{\textrm{o}}$, R.~P.~Sagar,
J.~M.~P$\acute{\textrm{e}}$rez-Jord$\acute{\textrm{a}}$,
V.~H.~Smith Jr., and R.~O.~Esquivel, Chem. Phys. Lett.
\textbf{219}, 15 (1994).

\bibitem{Ho2} M.~H$\hat{\textrm{o}}$, R.~P.~Sagar, D.~E.~Weaver, and V.~H.~Smith
Jr., Int. J. Quantum Chem., Quantum Chem. Symp. \textbf{29}, 109
(1995).

\bibitem{Ho3} M.~H$\hat{\textrm{o}}$, H.~Schmider, R.~P.~Sagar, D.~E.~Weaver, and
V.~H.~Smith Jr., Int. J. Quantum Chem. \textbf{53}, 627 (1995).

\bibitem{Ramirez98} J.~C.~Ramirez, J.~M.~Hern$\acute{\textrm{a}}$ndez
P$\acute{\textrm{e}}$rez, R.~P.~Sagar and R.~.~Esquivel, Phys.
Rev. A \textbf{58}, 3507 (1998).

\bibitem{Onicescu96} O.~Onicescu, R. Acad. Sci. Paris A
\textbf{263}, 25 (1996).

\bibitem{Moustakidis05} Ch.~C.~Moustakidis, K.~Ch.~Chatzisavvas,
and C.~P.~Panos, Int. J. Mod. Phys. E, (to be published), e-print
quant-ph/0506041.

\bibitem{Lepadatu04} C.~Lepadatu, and E.~Nitulescu, Acta.
Chim. Slov. \textbf{50}, 539 (2004).

\bibitem{Landsberg} P.~T.~Landsberg, Phys. Lett. A \textbf{102}, 171 (1984).

\bibitem{Kullback59} S.~Kullback, \emph{Statistics and Information
theory}, (Wiley, New York, 1959).

\bibitem{Wooters} A.~Majtey, P.~W.~Lamberti, M.~T.~Martin, and
A.~Plastino, e-print quant-ph/0408082.

\bibitem{Rao87} C.~Rao, \emph{Differential Geometry in Statistical
Interference}, IMS-Lecture Notes \textbf{10}, 217 (1987).

\bibitem {Lin91} J.~Lin, IEEE Trans. Inf. Theory \textbf{371},
145 (1991).

\bibitem{Sagar01}R.P. Sagar, J.C. Ramirez, R.O. Esquivel, M. Ho, and
V.H. Smith Jr., Phys. Rev. A {\bf 63}, 022509 (2001).

\bibitem{Sommerfeld32}A. Sommerfeld, Z. Phys. {\bf 78}, 283 (1932).

\bibitem{Guevara}N.L. Guevara, R.P. Sagar, and R.O. Esquivel,
J. Chem. Phys. {\bf 122}, 084101 (2005).

\bibitem{Gounaris} G.~J.~Gounaris, E.~A.~Paschos, and P.~I.~Porfyriadis,
Phys. Lett. B {\bf 525}, 63 (2002); G.~J.~Gounaris, E.~A.~Paschos,
and P.~I.~Porfyriadis, Phys. Rev. D {\bf 70}, 113008 (2004).

\bibitem{Vergados05} J.~D.~Vergados and H.~Ejiri, Phys. Lett. B \textbf{606}, 313
(2005).

\bibitem{Moustakidis}Ch.~C.~Moustakidis, H.~Ejiri and J.~D.~Vergados, e-print hep-ph/0507123.

\bibitem{Lopez95} R.~Lopez-Ruiz, H.~L.~Mancini, and X.~Calbet,
Phys. Lett. A \textbf{209}, 321 (1995).

\bibitem{Landsberg98} P.~T.~Landsberg and J.~Shiner, Phys. Lett. A \textbf{245}, 228 (1998).

\bibitem{Panos01c} C.~P.~Panos, Phys. Lett. A \textbf{289}, 287
(2001).

\end{thebibliography}
\end{document}